\documentclass[aps,prl,twocolumn,groupedaddress]{revtex4-2}
\usepackage{pifont}
\usepackage{amsmath}
\usepackage{rotating}
\usepackage{wasysym}
\usepackage{textcomp}
\usepackage{tikz}
\usepackage{pgfplots}
\usepgfplotslibrary{fillbetween}
\usetikzlibrary{patterns}
\usetikzlibrary{datavisualization}
\usetikzlibrary{arrows}
\usepgfplotslibrary{groupplots}
\usepackage{txfonts}

\begin{document}

\title{Influence of Iron Losses on Switching Dynamics\\ of an Electromagnet from Experiment and Simulation}


\author{Herbert Schmidt}
\email[]{corresponding author: herbert.schmidt@hs-bochum.de}
\author{Silvia Hacia}
\affiliation{Bochum University of Applied Sciences, D-42579 Heiligenhaus, Germany}


\date{January 24, 2020}

\begin{abstract}
The switching behaviour of electromagnets is determined by the inertia of the armature, the stiffness of the return spring and the magnetostatic forces between armature and yoke. For highly dynamic systems, hysteresis and eddy current losses have a slowing effect. In this paper we consider the experimentally observed behaviour of a switching magnet and compare it with simulation results including hysteresis and eddy current losses.

Paper presentend at the 12$^{\rm th}$ OpenModelica Annual Workshop in Link$\ddot{\rm o}$ping, Sweden, February 3, 2020.
\end{abstract}

\keywords{Equation-based modelling, Experimental verification, Magnetic circuits, Iron Losses, Hysteresis, Eddy Current}

\maketitle

\section{Considered Situation}\label{sec-problem}
Magnetic circuits can be described qualitatively and quantitatively in both geometry-based simulation (e.g. finite element method, FEM) and equation-based simulation (e.g. lumped elements modelling). In magnetostatics, geometry-based simulation is advantageous due to the fundamental freedom of the geometry under consideration. For dynamic calculations, equation-based simulation is advantageous, since it directly includes simultaneous coupling to surrounding systems. In this paper, we want to explore the capabilities of the OpenModelica Connection Editor (OMEdit) to accurately model highly dynamic switching operations.

\subsection{Soft Magnetic Material}\label{sec-material}

\begin{figure}[b]
\unitlength=0.01\linewidth
\begin{picture}(100,69.5)
\put(0,0){\begin{tikzpicture}
    \begin{axis}[
     width=1.0\linewidth,
     height=0.75\linewidth,
     xlabel={$H$ in [kA/m]},
     ylabel={$J$ in [T]},
     xlabel near ticks,
     ylabel near ticks,
     every tick/.style={color=black, thin},
     ticklabel style={/pgf/number format/precision=1,/pgf/number format/fixed,/pgf/number format/fixed zerofill,
     /pgfplots/ticklabel shift=0.005\linewidth,/pgfplots/major tick length=0.008\linewidth,/pgfplots/minor tick length=0.004\linewidth},
     minor x tick num=4,
     minor y tick num=9,
     xmin=-1.4, xmax=1.4,
     ymin=-1.3, ymax=1.3,
    ]
    \addplot[color=black,dashed,restrict x to domain=-2:2] table {1_4105_01.txt};
    \addplot[color=red,thin,restrict x to domain=-2:2] table {1_4105_02.txt};
     \end{axis}
\end{tikzpicture}}
\end{picture}
\caption{Initial magnetization curve and major hysteresis loop for X6CrMoS17. The initial magnetization curve is displayed as recorded (red solid line), the major hysteresis loop was subsequently symmetrized (black dashed line). This explains the minimal vertical offset.}
\label{fig-09072019-01}
\end{figure}
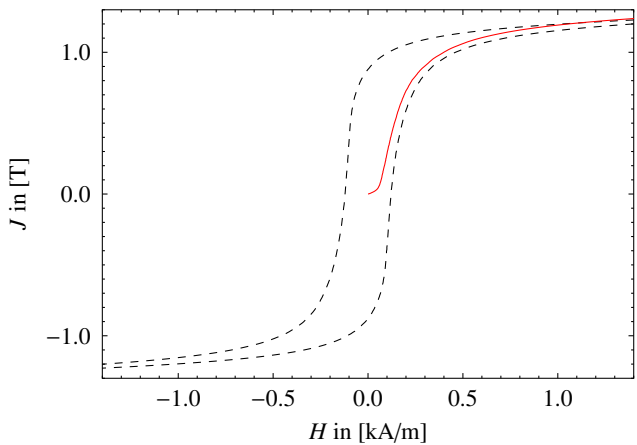

A simple flat armature switching magnet was considered, built from a ferritic stainless steel X6CrMoS17 (1.4105) for the flux-carrying parts, corresponding to ASTM A838, Alloy Type 2. 
 The initial magnetization curve and the major hysteresis loop, cf. Figure \ref{fig-09072019-01}, were tested up to 59.5~kA/m with a permeameter of type B according to IEC 60\,404--4 (Magnetphysik Dr. Steingr$\ddot{\text{o}}$ver ``Remagraph''). 
 A saturation polarization of 1.54~T, a maximum permeability of 2960, and a coercivity of 121~A/m were determined.

In equation-based calculations, an analytical representation of the permeability curve $\mu_r(B)$ is desired. $\mu_r$ denotes the relative permeability and $B$ the magnetic flux density:
\begin{equation*}
B=\mu_0\mu_rH=J+\mu_0H
\end{equation*}
where $\mu_0$ is the magnetic constant.
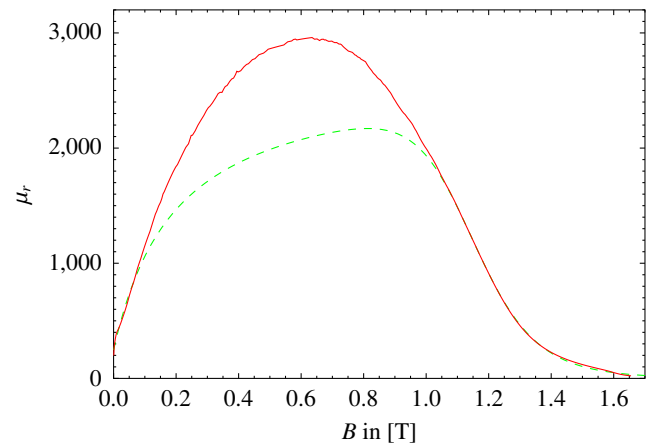
\begin{figure}[b]
\unitlength=0.01\linewidth
\begin{picture}(100,69.5)
\put(-1.5,0){
\begin{tikzpicture}
    \begin{axis}[
     width=1.0\linewidth,
     height=0.75\linewidth,
     xlabel={$B$ in [T]},
     ylabel={$\mu_r$},
     xlabel near ticks,
     ylabel near ticks,
     every tick/.style={color=black, thin},
     xticklabel style={/pgf/number format/precision=1,/pgf/number format/fixed,/pgf/number format/fixed zerofill,
     /pgfplots/ticklabel shift=0.005\linewidth,/pgfplots/major tick length=0.008\linewidth,/pgfplots/minor tick length=0.004\linewidth},
     yticklabel style={/pgf/number format/precision=0,/pgf/number format/fixed,/pgf/number format/fixed zerofill,
     /pgfplots/ticklabel shift=0.005\linewidth,/pgfplots/major tick length=0.008\linewidth,/pgfplots/minor tick length=0.004\linewidth},
     minor x tick num=3,
     minor y tick num=9,
     xmin=0, xmax=1.7,
     ymin=0, ymax=3200,
    ]
     \addplot[color=green,dashed,samples=100,domain=0:1.7,domain y=0:3000] {1+(246-1+13400*x/0.995)/(1+5*x/0.995+(x/0.995)^12.8)};
     \addplot[color=red,thin,samples=100,domain=0:1.7] table {14105muB.txt};
     \end{axis}
\end{tikzpicture}}
\end{picture}
\caption{Permeability of X6CrMoS17. Experimental data (red solid line), 
 fitting function $\hat{\mu}_r(B)$ (green dashed line).{\color{white}Permeabilit of X6CrMoS17. Exerimental data black solid line fittin function red dashed line. xxxxxxxxx Permeabilit of X6CrMoS17. Exerimental data black solid}}
\label{fig-09072019-2}
\end{figure}

OMEdit stores the initial magnetization curve of user-defined materials in the form of five fitting-parameters to the analytical function:
\begin{equation*}
\hat{\mu}_r(B)=1+\frac{\mu_i-1+c_aB_N}{1+c_bB_N+B_N^n}\label{Roschke-eqn4-16}
\quad
\text{where:}
\quad
B_N=\left|\frac{B}{B(\mu_\text{max})}\right|
\end{equation*}
Figure \ref{fig-09072019-2} shows the resulting function $\hat{\mu}_r(B)$ alongside the experimental data. 
 Note that the function $\hat{\mu}_r(B)$ has a characteristic shape that cannot be ideally matched to the measured data regardless of the parameter selection. The function shown was obtained by excluding the data for $\mu_r>1000$ from the fit. The obtained parameters are given in Table \ref{tab-09072019-1}.

The fact that these data were excluded can be justified as follows. Magnetic circuits effectively act as magnetic voltage dividers. As long as the permeance (magnetic conductivity) of one element is \textit{much} higher than that of other elements in the same circuit, changes in the exact value of the permeability will scarcely affect the effective behavior of the circuit. In comparison to the surrounding air with relative permeability of unity, it is expected that the resulting deviations for regions with permeability thousands of times higher anyway will have little effect on the overall result. In the range 1.02~T to 1.39~T the deviation from the data is in this case no more than $\pm2~\%$, for lower operating points the function $\hat{\mu}_r(B)$ falls systematically short of the actual permeability, cf. Figure \ref{fig-09072019-2}, as discussed just above.

This characteristic curve is used to perform magnetostatic calculations. In highly dynamic operation, hysteresis and eddy current losses become important, and the use of only the initial curve is no longer obviously justified. OMEdit offers a choice of two prepared hysteresis models, the Preisach model and the Tellinen model. For our purposes, the simpler Tellinen model will suffice \cite{Tellinen1998,Ziske2012}. 

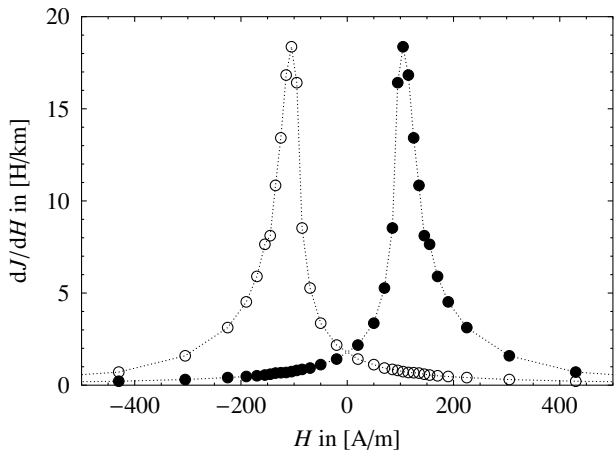
\begin{figure}
\unitlength=0.01\linewidth
\begin{picture}(100,69.5)
\put(-1.5,0){
\begin{tikzpicture}
    \begin{axis}[
     width=1.0\linewidth,
     height=0.75\linewidth,
     xlabel={$H$ in [A/m]},
     ylabel={${\rm d}J/{\rm d}H$ in [H/km]},
     xlabel near ticks,
     ylabel near ticks,
     every tick/.style={color=black, thin},
     xticklabel style={/pgf/number format/precision=0,/pgf/number format/fixed,/pgf/number format/fixed zerofill,
     /pgfplots/ticklabel shift=0.005\linewidth,/pgfplots/major tick length=0.008\linewidth,/pgfplots/minor tick length=0.004\linewidth},
     yticklabel style={/pgf/number format/precision=0,/pgf/number format/fixed,/pgf/number format/fixed zerofill,
     /pgfplots/ticklabel shift=0.005\linewidth,/pgfplots/major tick length=0.008\linewidth,/pgfplots/minor tick length=0.004\linewidth},
     minor x tick num=3,
     minor y tick num=4,
     xmin=-500, xmax=500,
     ymin=0, ymax=20,
    ]
    \addplot[color=black,densely dotted,restrict x to domain=-1000:1000] table[x index=0,y expr=\thisrowno{1}*1000] {12032019_Hysterese_abl.txt};
    \addplot[color=black,densely dotted,restrict x to domain=-1000:1000] table[x index=2,y expr=\thisrowno{3}*1000] {12032019_Hysterese_abl.txt};
    \addplot[color=black,only marks,restrict x to domain=-1000:1000,mark=o] table[x index=0,y expr=\thisrowno{1}*1000] {12032019_Hysterese_abl.txt};
    \addplot[color=black,only marks,restrict x to domain=-1000:1000,mark=*] table[x index=2,y expr=\thisrowno{3}*1000] {12032019_Hysterese_abl.txt};
     \end{axis}
\end{tikzpicture}}
\end{picture}
\caption{Hysteresis curve of X6CrMoS17, prepared for OMEdit. Derivative for the falling (open symbols) and rising branch (filled symbols). Symbols mark selected grid points.}
\label{fig-22112019-1}
\end{figure}

Only the major hysteresis loop $J(H)$ is needed, namely the upper limiting curve from saturation (falling branch) $J_-$ and the lower limiting curve (rising branch) $J_+$. Minor loops are reconstructed based on this information, the present value of $J$ and on whether the field strength is currently rising or falling:
\begin{equation*}
\frac{{\rm d}J}{{\rm d}H}=\left\{\begin{array}{ll}
\frac{J_--J}{J_--J_+}\frac{{\rm d}J_+}{{\rm d}H}&\text{if }{\rm d}H>0\\
\frac{J-J_+}{J_--J_+}\frac{{\rm d}J_-}{{\rm d}H}&\text{if }{\rm d}H<0\\
0&\text{else}
\end{array}\right.
\end{equation*}
Figure \ref{fig-22112019-1} shows the data points used in Modelica, forming the basis for the reconstruction. They were obtained by numerical derivation of the measured major hysteresis loop (cf. Figure \ref{fig-09072019-01}), symmetrization of the result and selection of 61 grid points at geometrically increasing distances from the maxima in the curves.

\begin{figure}
\unitlength=0.01\linewidth
\begin{picture}(100,69.5)
\put(-1.5,0){
\begin{tikzpicture}
    \begin{axis}[
     width=1.0\linewidth,
     height=0.75\linewidth,
     xlabel={$H$ in [A/m]},
     ylabel={$J$ in [T]},
     xlabel near ticks,
     ylabel near ticks,
     every tick/.style={color=black, thin},
     xticklabel style={/pgf/number format/precision=0,/pgf/number format/fixed,/pgf/number format/fixed zerofill,
     /pgfplots/ticklabel shift=0.005\linewidth,/pgfplots/major tick length=0.008\linewidth,/pgfplots/minor tick length=0.004\linewidth},
     yticklabel style={/pgf/number format/precision=1,/pgf/number format/fixed,/pgf/number format/fixed zerofill,
     /pgfplots/ticklabel shift=0.005\linewidth,/pgfplots/major tick length=0.008\linewidth,/pgfplots/minor tick length=0.004\linewidth},
     minor x tick num=4,
     minor y tick num=4,
     xmin=0, xmax=500,
     ymin=0, ymax=1.1,
    ]
     \addplot[color=red,thin,restrict x to domain=0:1000] table [x index=1, y index=3]{1_4105_HBJ_Am.txt};
	\addplot[color=black,densely dotted,restrict x to domain=0:600] table [x index=0, y index=2]{12032019_Hysterese_JH_01mum.txt};
    \addplot[color=green,dashed,restrict x to domain=0:1000] table{1_4105_neu_OMEdit.txt};
    \end{axis}
\end{tikzpicture}}
\end{picture}
\caption{Initial curve of X6CrMoS17, shown are the measured curve (red solid line), the curve interpolated using the Tellinen model corresponding to Figure \ref{fig-22112019-1} (black dotted line), and the curve used by OMEdit without consideration of hysteresis corresponding to Figure \ref{fig-09072019-2} (green dashed line).}
\label{fig-22112019-2}
\end{figure}
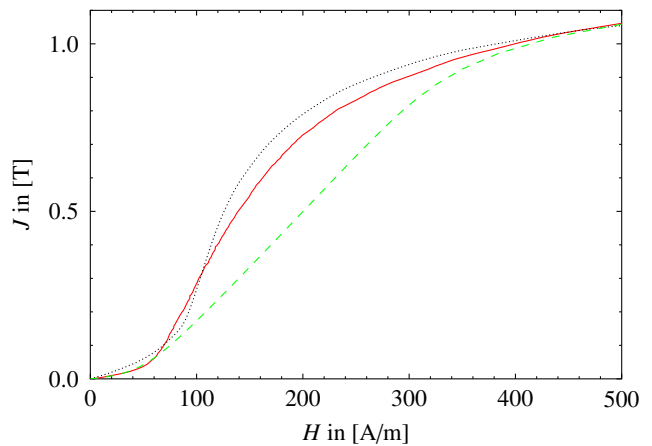

Figure \ref{fig-22112019-2} exemplarily shows the reconstruction of the initial curve using the Tellinen model (black dotted line) together with the measured initial curve (red solid line). Both are in reasonably good, yet not perfect agreement (due to the simplified structure of the Tellinen model). For comparison, note that the initial curve resulting from $B(\mu)$ as shown in Figure \ref{fig-09072019-2} (green dashed line), deviates significantly further from the measured initial curve over the identical value range.

In order to include eddy currents, we need a resistivity reading for the X6CrMoS17 used. A four-terminal measurement was performed using the cylindrical sample from the permeameter measurement. The voltage sensor tips were separated by 65.0~mm, the sample had a cross-sectional area of 113~mm$^2$ ($\pm0.25$~\% relative uncertainty for either). For our simulations we will use the resulting nominal resistivity of 0.774~\textmu$\Omega$\,m ($\pm0.35$~\% relative uncertainty).

\begin{table}
\caption{Permeability of X6CrMoS17. Parameters of fitting function in OMEdit.}
\label{tab-09072019-1}
\begin{tabular}{ccccc}
\hline\hline
$\mu_i$&$B(\mu_\text{max})$&$c_a$&$c_b$&$n$\\
246&0.995~T&13,400&5&12.8\\
\hline\hline
\end{tabular}
\end{table}

\subsection{Geometry}

\begin{figure}
\unitlength=0.01\linewidth
\begin{picture}(100,44.3)
\put(0,0){
\put(68.6,0){\includegraphics[scale=0.5]{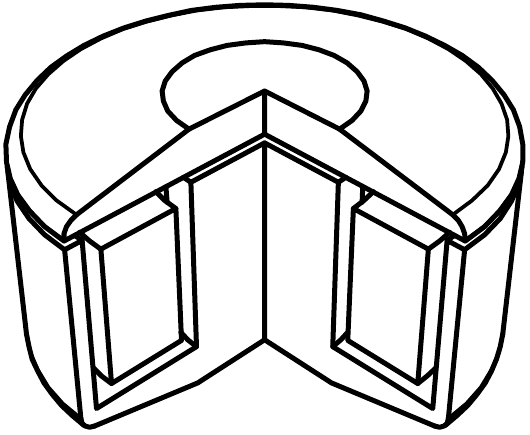}}
\put(-25,0){\put(25,0.05){\includegraphics[width=0.675\linewidth]{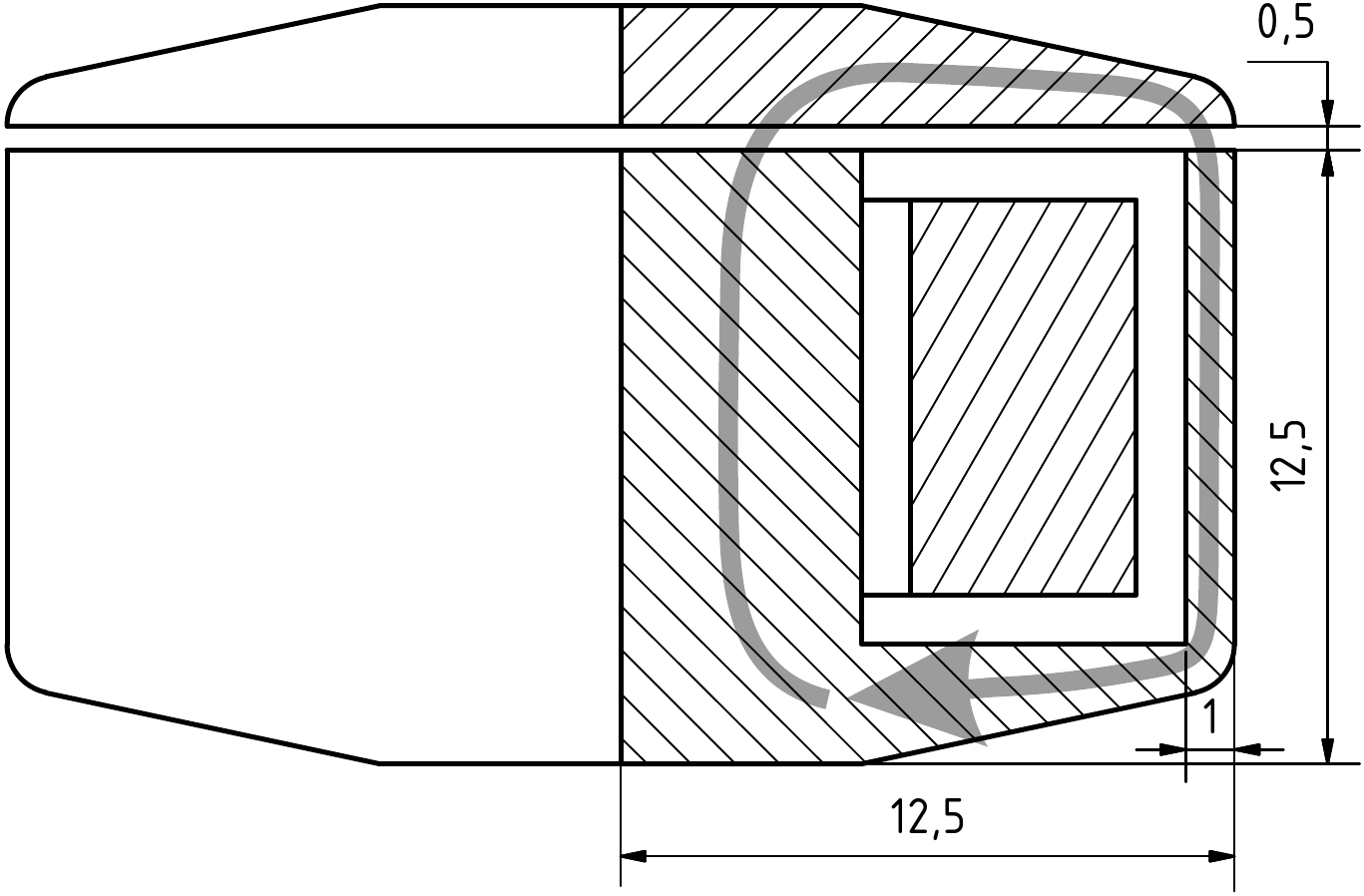}}
\put(0,-5){
\put(93.7,41.5){air gap}
\put(33,44){armature}
\put(54,33){\color{white}\rule{0.075\linewidth}{0.003\linewidth}}
\put(54,33.15){\vector(1,0){7.5}}
\put(37,32){inner pole}
\put(72.2,27.5){\color{white}\rule{0.065\linewidth}{0.038\linewidth}}
\put(72.5,28){coil}
\put(33,28){yoke}
\put(54,25){\color{white}\rule{0.31\linewidth}{0.003\linewidth}}
\put(54,25.15){\vector(1,0){31}}
\put(37,24){outer pole}
\put(45,15){\color{white}\rule{0.27\linewidth}{0.003\linewidth}}
\put(45,15.15){\vector(1,0){27}}
\put(37,14){base}
}
}
}
\end{picture}
\caption{Simple switching magnet.}
\label{fig-10072019-1}
\end{figure}
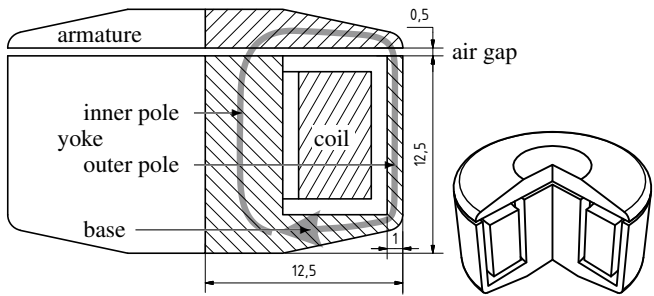

We consider a switching magnet, composed of a yoke with embedded coil and a flat armature, cf. Figure \ref{fig-10072019-1}. In this figure a single flux vortex is symbolically represented. The diameter is \diameter 25~mm, the height of the yoke is half that, 12.5~mm. The nominal air gap is 0.5~mm. Since the considered vortex traverses this gap twice, an air gap of 1.0~mm is effectively realized.  We assume a current linkage of 800~A and select all other geometries in such a way that an essentially constant cross-section is realized throughout the circuit. This results in a nominally homogeneous working point, facilitating quantitatively accurate results in equation-based simulation.

\begin{figure}[b]
\unitlength=0.01\linewidth
\begin{picture}(100,76.1)
\put(42.925,0){\includegraphics[width=0.57075\linewidth]{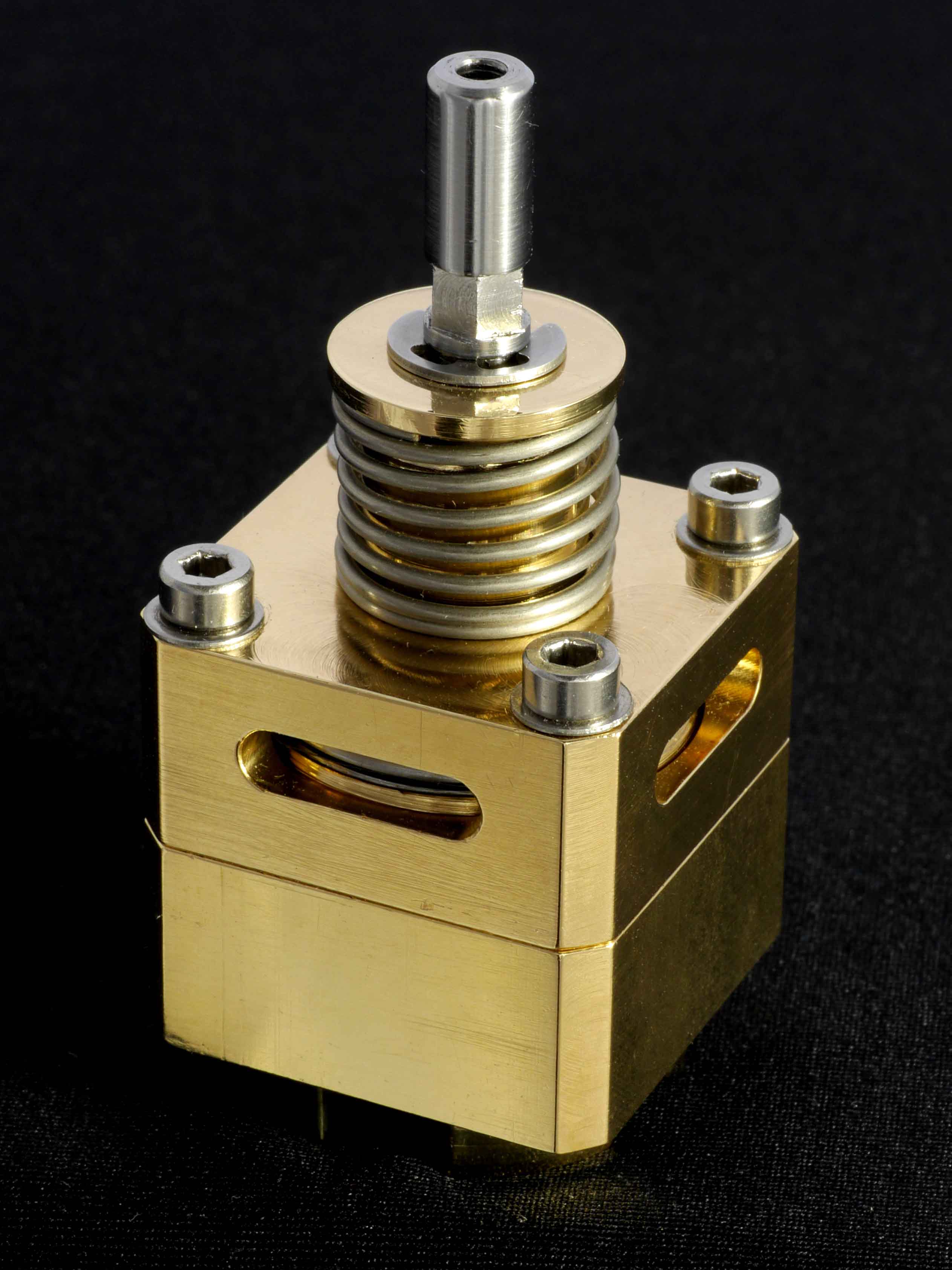}}
\put(0,0){\includegraphics[scale=0.5]{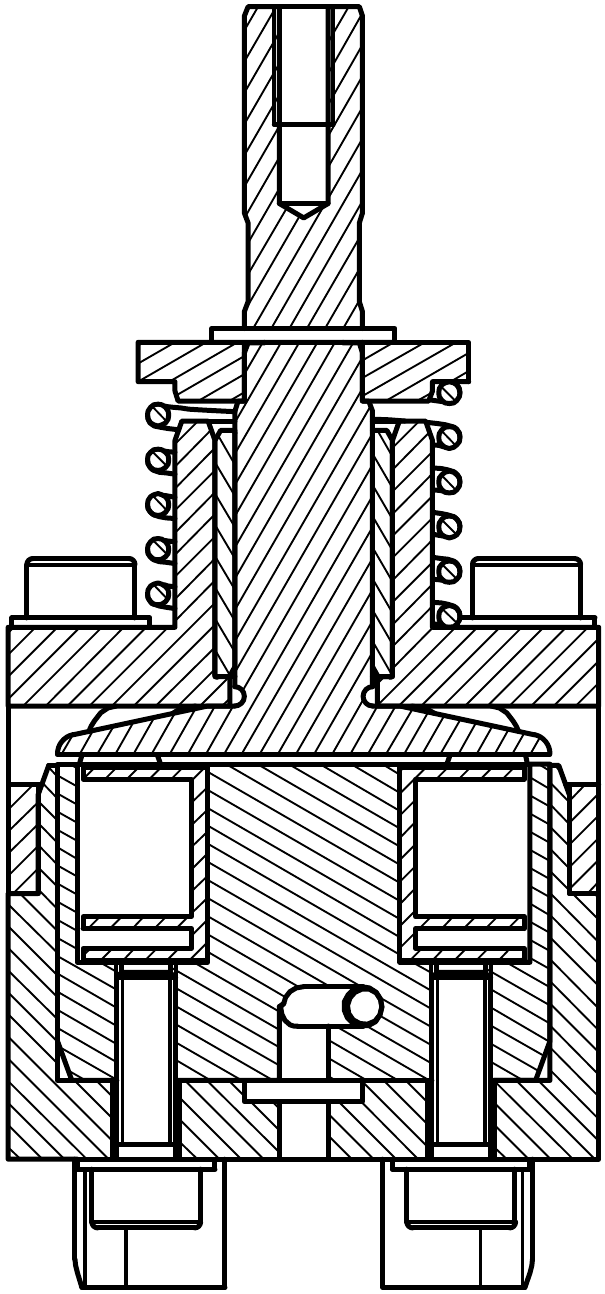}}
\end{picture}
\caption{Fully assembled functional model, original size.}
\label{fig-10072019-2}
\end{figure}

For the selected material X6CrMoS17, a cross-sectional area (perpendicular to the direction of magnetic flux) of 75.4~mm$^2$ (corresponding to 1~mm wall thickness in the outer pole) results in a working point of 1.0~T with a relative permeability of 1996. Correspondingly, the inner pole has a diameter of 9.8~mm and the armature has a thickness of 2.45~mm.

For the functional model, the geometry 
 had to be 
 adapted to enable guides, brackets, stops, etc., cf. Figure \ref{fig-10072019-2}. Particularly close to the working air gap, care was taken not to alter the geometries of the ferromagnetic components, armature and yoke. All components 
 apart from the nominal magnetic circuit were manufactured from non-magnetic materials. 
  The return spring was mounted above the magnetic circuit and acts against a plate 
   fixed to the armature needle with a spring washer. 
   
The housing was designed in two parts so that the magnet can be disassembled along the working air gap by loosening four screws without loosening the return spring. In the upper part of the housing, transverse inspection openings have been provided so that the movement of the armature can be observed directly.
 The armature needle was extended upwards in order to be able to couple to force or displacement measurements. The lower part of the housing has corresponding connection geometries. The yoke is fixed from below by two screws, the bobbin is pinned into the yoke. The connecting wires are extracted axially downwards. 
The current linkage of 800~A is thus realized in a window of $(7.0\times 5.5)$~mm$^2=38.5\text{ mm}^2$. This results in a current density of 21~A/mm$^2$. Please note that this is a switching magnet and no continuous operation at this level of DC current density is required.
   The coil was made from 131 turns of \diameter 0.45~mm enamelled copper wire, with a nominal DC resistance of 0.75~$\Omega$. Then 6.1~A DC are required for the assumed 800~A current linkage. Experimentally, 8.0~A were available.

\subsection{Dynamic testing}

\begin{figure}[b]
\includegraphics[width=\linewidth]{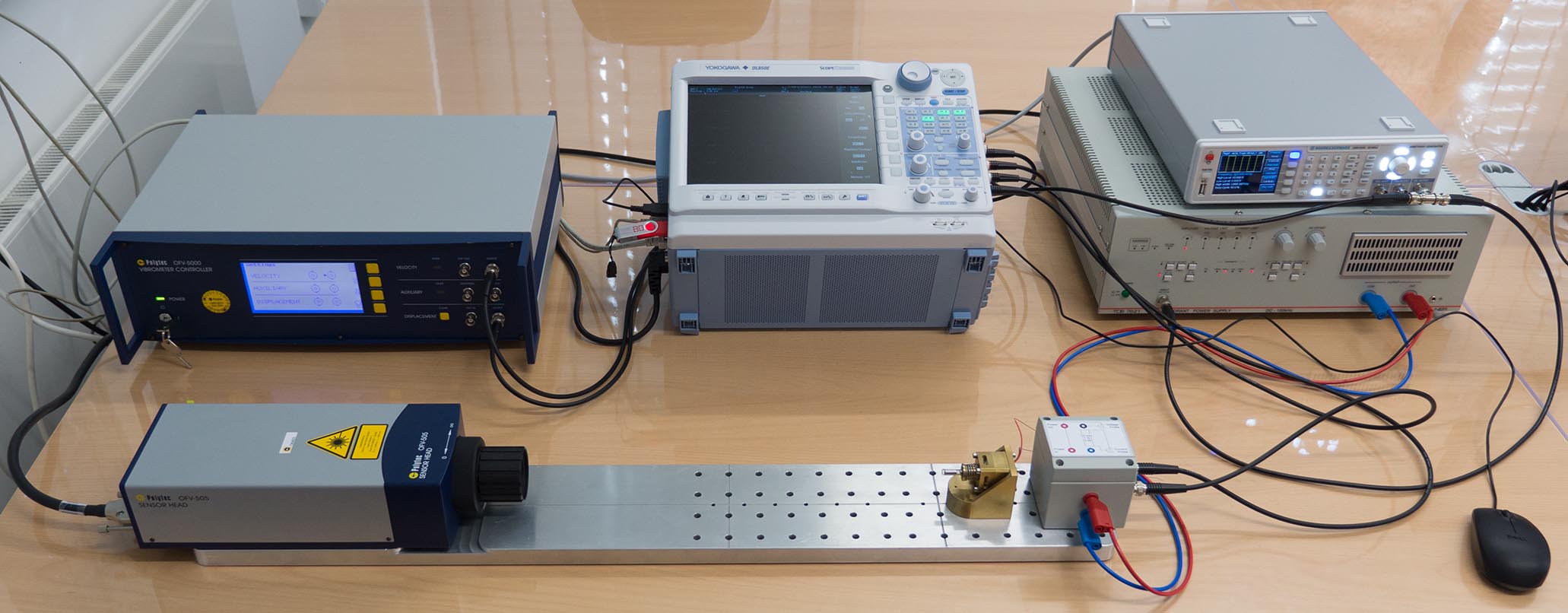}
\vskip2mm
\centerline{\includegraphics[width=\linewidth]{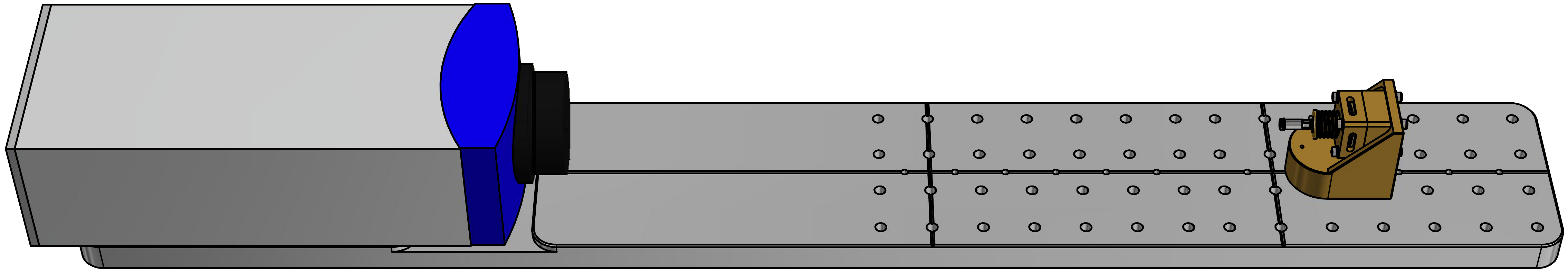}}
\caption{Laser test rig for dynamical measurements.}
\label{fig-11102018-01}
\end{figure}

In preparation for dynamic measurements, the moving masses were weighed, including half of the mass of the return spring. The resulting value of 19.9~g was used for dynamic simulations.

The stiffness of the return spring was measured using a tensile testing machine. Over the used stroke range the spring rate is 0.7854~N/mm ($\pm0.34$~\% relative uncertainty). This value corresponds to a calculated unstressed spring length of 39.51~mm (12.0~mm at open air gap, 12.5~mm at closed air gap) in dynamic simulations.

The electromagnet is controlled by a Rohde \& Schwarz HMF 2550 function generator whose (preset) output voltage signal is converted into a corresponding (actual) current-controlled signal by a Toellner TOE 7621 four-quadrant amplifier. Both the preset and the actual current (measured via a 10~m$\Omega$ shunt) are recorded, as well as the resulting voltage over the electromagnet. Displacement and velocity information is measured by a Polytec OFV 5000 (controller) and OFV 505 (optics) laser vibrometer. A reflector is screwed onto the armature. Laser and electromagnet are each fixed to a common base plate in such a way that the reflection takes place at (the second) maximum visibility of the laser (cf. Figure \ref{fig-11102018-01}).

\begin{figure}
\unitlength=0.01\linewidth
\begin{picture}(100,69.5)
\put(-1.5,0){
\begin{tikzpicture}
\begin{groupplot}[group style={group size=1 by 2,
		    xlabels at=edge bottom,
               xticklabels at=edge bottom,
               yticklabels at=edge left,
               vertical sep=0.01\linewidth,
               horizontal sep=0.01\linewidth},
               height=0.4635\linewidth,width=1\linewidth]
\nextgroupplot[
      xticklabel style={/pgf/number format/precision=0,/pgf/number format/fixed,/pgf/number format/fixed zerofill,
     /pgfplots/ticklabel shift=0.005\linewidth,/pgfplots/major tick length=0.008\linewidth,/pgfplots/minor tick length=0.004\linewidth},
     yticklabel style={/pgf/number format/precision=1,/pgf/number format/fixed,/pgf/number format/fixed zerofill,
     /pgfplots/ticklabel shift=0.005\linewidth,/pgfplots/major tick length=0.008\linewidth,/pgfplots/minor tick length=0.004\linewidth},
ylabel={$d_\text{ag}$ in [mm]},
     minor x tick num=9,
     minor y tick num=3,
     xmin=0, xmax=5,
     ymin=-0.05, ymax=0.55,
]
	\addplot[color=green,thin,samples=100] table[x expr=\thisrowno{0}-10,y index=2]{Schalten_80_zu_kurz2.txt}; 
	\addplot[color=green,thin,samples=100] table[x index=0,y index=2]{Schalten_80_auf_kurz.txt}; 
	\addplot[color=red,dashed,samples=100] table[x index=0,y index=1]{Modelica_Schaltvergleich2.txt}; 
\nextgroupplot[
     xticklabel style={/pgf/number format/precision=0,/pgf/number format/fixed,/pgf/number format/fixed zerofill,
     /pgfplots/ticklabel shift=0.005\linewidth,/pgfplots/major tick length=0.008\linewidth,/pgfplots/minor tick length=0.004\linewidth},
     yticklabel style={/pgf/number format/precision=0,/pgf/number format/fixed,/pgf/number format/fixed zerofill,
     /pgfplots/ticklabel shift=0.005\linewidth,/pgfplots/major tick length=0.008\linewidth,/pgfplots/minor tick length=0.004\linewidth},
     ylabel={$v$ in [m/s]},
     xlabel={$t$ in [ms]},
     minor x tick num=9,
     minor y tick num=4,
     xmin=0, xmax=5,
     ymin=-2.0, ymax=1.2,
]
	\addplot[color=green,thin,samples=100] table[x expr=\thisrowno{0}-10,y index=1]{Schalten_80_zu_kurz2.txt}; 
	\addplot[color=green,thin,samples=100] table[x index=0,y index=1]{Schalten_80_auf_kurz.txt}; 
	\addplot[color=red,dashed,samples=100] table[x index=0,y index=2]{Modelica_Schaltvergleich2.txt}; 
\end{groupplot}
\end{tikzpicture}}
\end{picture}
\caption{Measured and simulated dynamics using 8.0~A of driving current and a nominally square wave form. Green solid line: measurement; red dashed line: simulation. 
 Top: armature stroke, bottom: armature speed.}
\label{fig-26112019-01}
\end{figure}
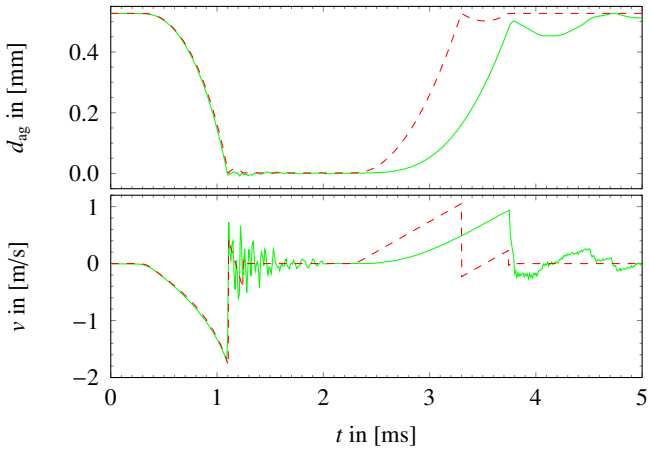

Figure \ref{fig-26112019-01} shows an example of such a dynamic measurement, in direct comparison with an OMEdit simulation (the models are discussed below). The time offset is chosen such that at $t=0$ the current is switched on and at $t=2$~ms the current is switched off. Obviously the closing process of the air gap is reproduced very well, while the opening process is well reproduced only as far as the shape of the curve is concerned, but takes place 0.451~ms later than simulated. In this article we discuss ways to quantitatively understand and model this delay.

\section{Modelling}
The procedure for network modelling of magnetic circuits has already been described in detail in the past \cite{Roters, Kallenbach, Schmidt}. The magnetostatic model correspondig to this exact switching magnet has been described in \cite{SchmidtEOOLT19}. We therefore concentrate on some aspects that are specific to the dynamic model chosen here and to modelling iron losses. All classes taken from the Modelica Standard Library are simply given with their path in \textit{italics}.

\begin{figure}
\unitlength=0.01\linewidth
\includegraphics[width=\linewidth]{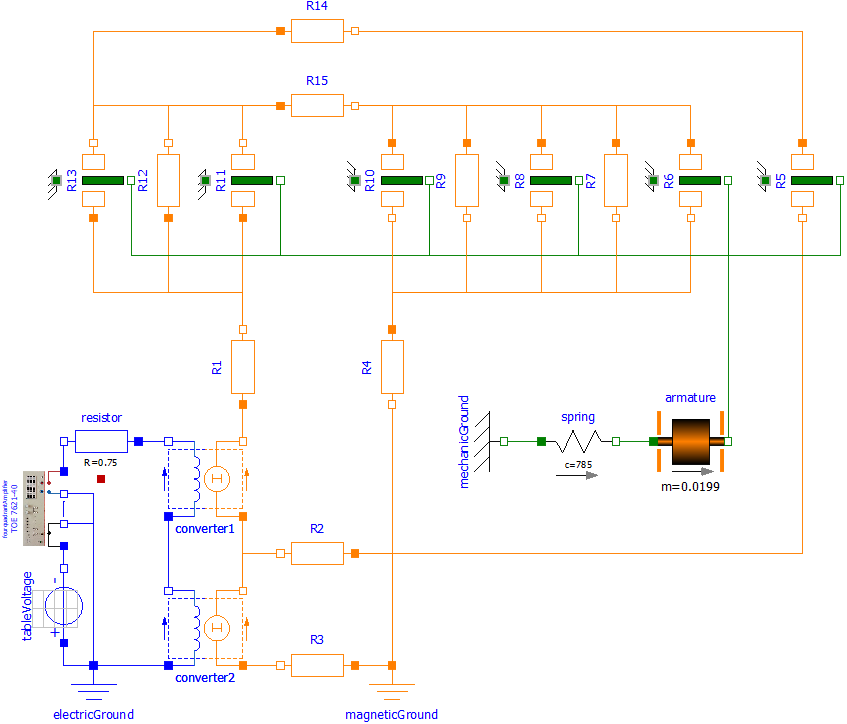}
\caption{Network model without iron losses in the Diagram View of OMEdit.}
\label{fig-17072019-1}
\end{figure}

\subsection{Model without iron losses}
The \textbf{magnetic submodel} (shown in orange in Figure \ref{fig-17072019-1}) initially remains unchanged compared to the magnetostatic case \cite{SchmidtEOOLT19}. It is based on Finite Element Method analysis of the situation at 0.5~mm air gap and 800~A current linkage. All iron elements use the initial curve of X6CrMoS17 according to $\hat{\mu}_r$ described above and actual geometries. One stray flux term accross the coil window was included as well as one from the outer surface of the yoke to the upper surface of the armature. Out of the terms that together constitute the working flux over the air gap between the yoke and the armature, those are linked to the mechanical system whose peremance changes with the air gap (these are the ones that contribute to the magnetostatic force).

In the \textbf{mechanical submodel} (shown in green), the combination of an armature mass and a fixed position preset used in the magnetostatic case was replaced by \textit{Modelica.\-Magnetic.\-Flux\-Tubes.\-Examples.\-Utilities.\-Translatory\-Armature\-And\-Stopper} for the ar\-ma\-ture-and-stops assembly, \textit{Modelica.\-Mechanics.\-Trans\-la\-tio\-nal.\-Components.\-Spring} for the return spring and \textit{Mo\-de\-li\-ca.\-Mecha\-nics.\-Trans\-la\-ti\-o\-nal.\-Com\-po\-nents.\-Fixed} for the support. For the spring, the measured spring rate of 785 ~N/m was used along with an unstressed spring length of 0.027\,01~m (12.5~mm actual length at closed airgap was substracted from the previously quoted value of 39.51~mm, i.e. we focus on the stroke alone, not the absolute dimension of this element). For the armature and stopper, we again disregard the absolute dimension and set the length to $L$\,=\,0. The armature mass corresponds to the measured value of $m$\,=\,0.0199~kg. The position of stopper is set to $x_\text{max}$\,=\,0.000\,527~m and $x_\text{min}$\,=\,0.000\,001~m. The maximum corresponds to the actual realized stroke, the minimum residual air gap of 1~\textmu m was chosen to avoid numerical problems. The parametrization of the impact at the respective stoppers was heuristically chosen such that the simulation is qualitatively in agreement with typical measurements, $c$\,=\,$10^{17}$~N/m, $d$\,=$\,10^{5}$~N\,s/m and $n$\,=\,2.

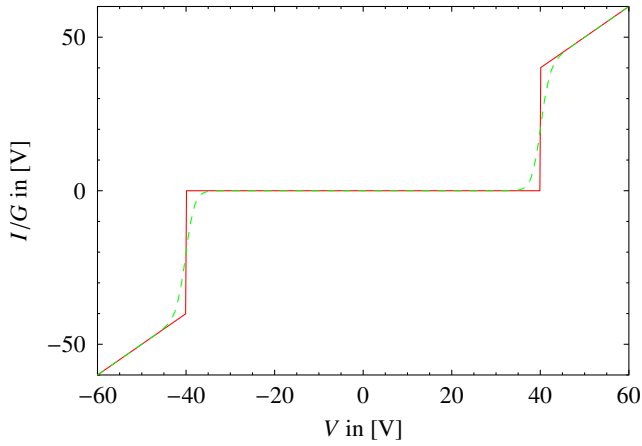
\begin{figure}
\unitlength=0.01\linewidth
\begin{picture}(100,69.5)
\put(0,0){\begin{tikzpicture}
    \begin{axis}[
     width=1.0\linewidth,
     height=0.75\linewidth,
     xlabel={$V$ in [V]},
     ylabel={$I/G$ in [V]},
     xlabel near ticks,
     ylabel near ticks,
     every tick/.style={color=black, thin},
     xticklabel style={/pgf/number format/precision=0,/pgf/number format/fixed,/pgf/number format/fixed zerofill,
     /pgfplots/ticklabel shift=0.005\linewidth,/pgfplots/major tick length=0.008\linewidth,/pgfplots/minor tick length=0.004\linewidth},
     yticklabel style={/pgf/number format/precision=0,/pgf/number format/fixed,/pgf/number format/fixed zerofill,
     /pgfplots/ticklabel shift=0.005\linewidth,/pgfplots/major tick length=0.008\linewidth,/pgfplots/minor tick length=0.004\linewidth},
     minor x tick num=3,
     minor y tick num=4,
     xmin=-60, xmax=60,
     ymin=-60, ymax=60,
    ]
    \addplot[color=red,thin] table[x index = {0}, y index = {1}] {spannungsbegrenzung.txt};
    \addplot[color=green,dashed] table[x index = {0}, y index = {2}] {spannungsbegrenzung.txt};
    \end{axis}
\end{tikzpicture}}
\end{picture}
\caption{Analytical description of the voltage limitation for $V_0$\,=\,40~V. Green dashed line: $d_V$\,=\,1~V and $b$\,=\,40 (the right and left sigmoid functions meet in the origin continuously differentiable). Red solid line: $d_V$\,=\,0.01~V and $b$\,=\,20 (used below).}
\label{fig-17072019-2}
\end{figure}

Figure \ref{fig-17072019-1} shows the \textbf{electrical submodel} in blue.
While for the magnetostatic simulation the power supply could simply be represented as an ideal current source with additional parallel resistance \cite{SchmidtEOOLT19}, this simplification for the dynamic simulation was replaced by a circuit similar to the actual technical situation. The signal generator is represented by \textit{Modelica.Electrical.Analog.Sources.TableVoltage}, where we use a square wave function as we did on the actual instrument. An extremely simplified model for the four-quadrant amplifier was generated from an ideal voltage-controlled current source, \textit{Modelica.Electrical.Spice3.Basic.G\_VCC}, whose output side is short-circuited by a voltage limitation. This voltage limitation is realized by a voltage-dependent resistor:  
\begin{equation*}
\frac{I}{G}=\left\{\begin{array}{lll}	
\frac{V}{1+\exp (-b)}&\text{if:}&\hskip19.2mmV\le-V_0-bd_V\\
\frac{V}{1+\exp\left((V_0+V)/d_V\right)}&\text{if:}&-V_0-bd_V< V< -V_0+bd_V\\
\frac{V}{1+\exp (+b)}&\text{if:}&-V_0+bd_V\le V\le{\color{white}-} V_0-bd_V\\
\frac{V}{1+\exp\left((V_0-V)/d_V\right)}&\text{if:}&{\color{white}-}V_0-bd_V< V<{\color{white}-} V_0+bd_V\\
\frac{V}{1+\exp (-b)}&\text{if:}&{\color{white}-}V_0+bd_V\le V
\end{array}\right.
\end{equation*}
with $V$ the voltage, $I$ the current, and $G=10$~kS the conductance outside the limiting voltage of $V_0=40$~V. Two additional parameters describe the width of the transition and the spread between conductive and blocking behavior, $d_V=0.01$~V and $b=20$. These parameters correspond to a change of the conductance by a factor of $4\cdot10^8$ over a window of 0.2~V, cf. Figure \ref{fig-17072019-2}. This function is continuous, but not differentiable in the exact sense. For the selected parameters, the first derivatives are sufficiently smooth for the intended use.

\subsection{Model including hysteresis losses}
\begin{figure}
\includegraphics[width=0.865\linewidth]{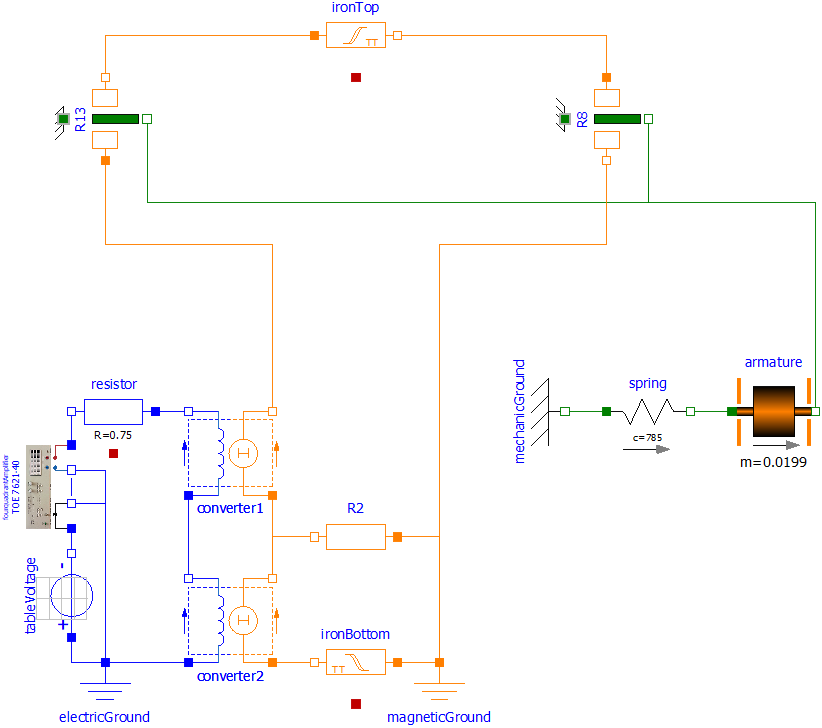}
\caption{Network model of the switching magnet including magnetic hysteresis, neglecting stray flux. Spatial arrangement of the remaining elements is identical to that in Figure \ref{fig-17072019-1}.}
\label{fig-08032019-01}
\end{figure}

The previously observed delay in switch-off could in principle be caused by remanent magnetostatic force. In fact, we will see that the coercivity in our soft magnetic X6CrMoS17 is far too low to have a significant effect, but we still want to investigate quantitatively the effect of magnetic hysteresis on dynamics. We use a 
 simplified network in which all flux leakage paths, except one across the coil window, are suppressed (cf. Figure \ref{fig-08032019-01}). The reluctances in yoke and armature are replaced using the class \textit{Mo\-de\-li\-ca.\-Mag\-ne\-tic.\-Flux\-Tubes.\-Shapes.\-Hys\-te\-re\-sis\-And\-Mag\-nets.\-Ge\-ne\-ric\-Hyst\-Tel\-li\-nen\-Table}, ini\-ti\-a\-li\-zed as depicted in Figure \ref{fig-09072019-2}. 

\begin{figure}[b]
\unitlength=0.01\linewidth
\begin{picture}(100,69.5)
\put(0,0){\begin{tikzpicture}
    \begin{axis}[
     width=1.0\linewidth,
     height=0.75\linewidth,
     xlabel={$H$ in [A/m]},
     ylabel={$B$ in [T]},
     xlabel near ticks,
     ylabel near ticks,
     every tick/.style={color=black, thin},
     xticklabel style={/pgf/number format/precision=0,/pgf/number format/fixed,/pgf/number format/fixed zerofill,
     /pgfplots/ticklabel shift=0.005\linewidth,/pgfplots/major tick length=0.008\linewidth,/pgfplots/minor tick length=0.004\linewidth},
     yticklabel style={/pgf/number format/precision=1,/pgf/number format/fixed,/pgf/number format/fixed zerofill,
     /pgfplots/ticklabel shift=0.005\linewidth,/pgfplots/major tick length=0.008\linewidth,/pgfplots/minor tick length=0.004\linewidth},
     minor x tick num=4,
     minor y tick num=4,
     xmin=-150, xmax=150,
     ymin=0, ymax=1.1,
    ]
	\addplot[color=black,thin] table {08032019_BH_3480nm.txt}; 
	\addplot[color=red,dashed] table {08032019_BH_3000nm.txt}; 
	\addplot[color=blue,dotted] table {08032019_BH_2000nm.txt}; 
	\addplot[color=green,densely dashed] table {08032019_BH_1000nm.txt}; 
	\addplot[color=brown,densely dotted] table {08032019_BH_100nm.txt}; 
    \addplot[color=black,thin] coordinates {(0,0)(-150,3.692628)};
    \addplot[color=black,thin] coordinates {(0,0)(-150,1.846314)};
    \addplot[color=black,thin] coordinates {(0,0)(-150,0.923157)};
    \end{axis}
\end{tikzpicture}}
\put(34,63){1 \textmu m}
\put(22.5,63){2 \textmu m}
\put(16,43){4 \textmu m}
\end{picture}
\caption{Influence of the residual air gap. Hysteresis curves for different residual air gaps that prevent the magnet from opening again. From outside in: 3.48 (black solid line), 3 (red dashed line), 2 (blue dotted line), 1 (green densely dashed line) and 0.1~\textmu m (brown densely dotted line). 
 In addition, three load lines are shown for residual air gaps as marked.}
\label{fig-28112019-1}
\end{figure}
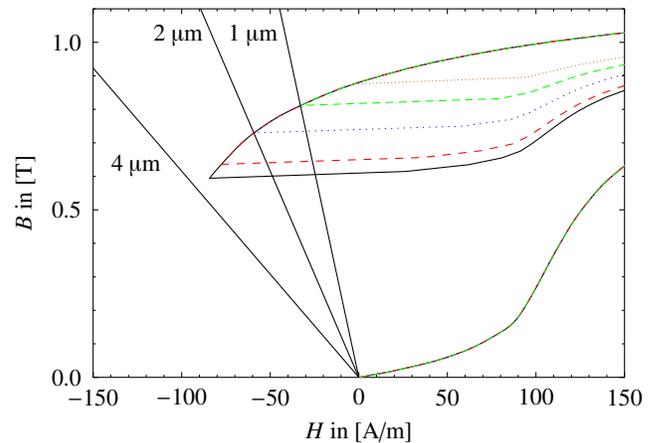

We now select different residual air gaps between 0.1 and 10~\textmu m in \textit{TranlatoryArmatureAndStopper} and simulate the switching behavior. In Figure \ref{fig-28112019-1}, the $B(H)$ characteristics are given for residual air gaps between 3.480 and 0.100~\textmu m with cyclic applied current (square wave form). The straight lines represent load lines for the specified residual air gaps. As expected, the corresponding characteristic curves in the second quadrant are reversed at these load lines. Only when the flux density in the residual air gap drops sufficiently that the magnetostatic force can be overcome by the return spring does the magnet open again. This is not the case for the characteristic curves shown in Figure \ref{fig-28112019-1}.

\begin{figure}
\unitlength=0.01\linewidth
\begin{picture}(100,69.5)
\put(0,0){\begin{tikzpicture}
    \begin{axis}[
     width=1.0\linewidth,
     height=0.75\linewidth,
     xlabel={$t$ in [ms]},
     ylabel={$v$ in [m/s]},
     xlabel near ticks,
     ylabel near ticks,
     every tick/.style={color=black, thin},
     xticklabel style={/pgf/number format/precision=1,/pgf/number format/fixed,/pgf/number format/fixed zerofill,
     /pgfplots/ticklabel shift=0.005\linewidth,/pgfplots/major tick length=0.008\linewidth,/pgfplots/minor tick length=0.004\linewidth},
     yticklabel style={/pgf/number format/precision=1,/pgf/number format/fixed,/pgf/number format/fixed zerofill,
     /pgfplots/ticklabel shift=0.005\linewidth,/pgfplots/major tick length=0.008\linewidth,/pgfplots/minor tick length=0.004\linewidth},
     minor x tick num=4,
     minor y tick num=4,
     xmin=2, xmax=5,
     ymin=-0.3, ymax=1.1,
         ]
	\addplot[color=black,thin] table {08032019_10000nm.txt}; 
	\addplot[color=red,dashed] table {08032019_4000nm.txt}; 
	\addplot[color=blue,dotted] table {08032019_3485nm.txt}; 
	\addplot[color=green,densely dashed] table {08032019_3482nm.txt}; 
	\addplot[color=brown,densely dotted] table {08032019_3481nm.txt}; 
    \end{axis}
\end{tikzpicture}}
\end{picture}
\caption{Influence of the residual air gap. Speed curves upon opening for different residual air gaps that do open again. From left to right: 10 (black solid line), 4 (red dashed line), 3.485 (blue dotted line), 3.482 (green densely dashed line) and 3.481~\textmu m (brown densely dotted line).}
\label{fig-28112019-2}
\end{figure}
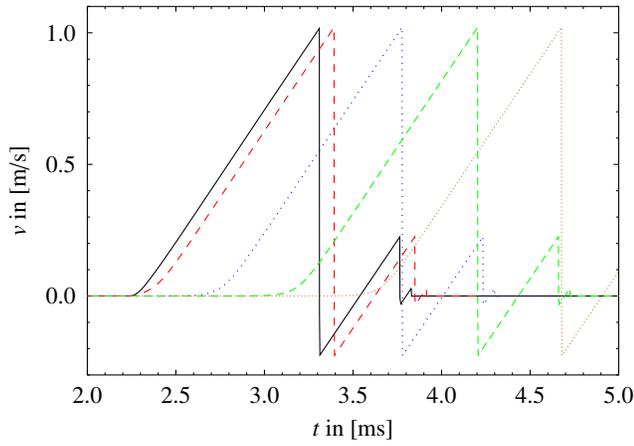

If we increase the residual air gap by just one more nanometer to 3.481~\textmu m, the magnet opens (in the simulation these length scales work), but with a delay of 1.367~ms compared to e.g.\ 10~\textmu m residual air gap (cf.\ Figure \ref{fig-28112019-2}). For intermediate values any delay can be realized, including the 0.451~ms which we noticed in Figure \ref{fig-26112019-01}. Nevertheless, we know that this is not the root cause for the experimentally observed delay.

This sensitive dependence on the exact value of the residual air gap is one reason why we only considered a very simplified magnetic network here. By selecting an air gap on the nanometer level, essentially any results can be achieved. A more complex model would not change this observation. At the same time, this sensitivity contradicts the observed robustness of the experimental effect. In addition, the numerical values in the range of 3 to 4~\textmu m contradict the experimental observation in the magnetostatic experiment \cite{SchmidtEOOLT19} that an effective residual air in the range of 50~\textmu m remains in the system even for mechanically fully closed air gap. And finally, a comparison of the shape of the curves between Figure \ref{fig-26112019-01} and \ref{fig-28112019-2} shows that in both cases a delay is realized, but that in Figure \ref{fig-26112019-01} the shape of the curve changes (it is rounded off), but not in Figure \ref{fig-28112019-2}. Since the spatial range in which remanent magnetization plays a role here is extremely small, the majority of the motion takes place according to exactly the same dynamics as without remanence. In the actual experiment we by contrast see an influence, which shows effect over a larger range of motion.

\subsection{Model including eddy current losses}
\begin{figure}
\includegraphics[width=\linewidth]{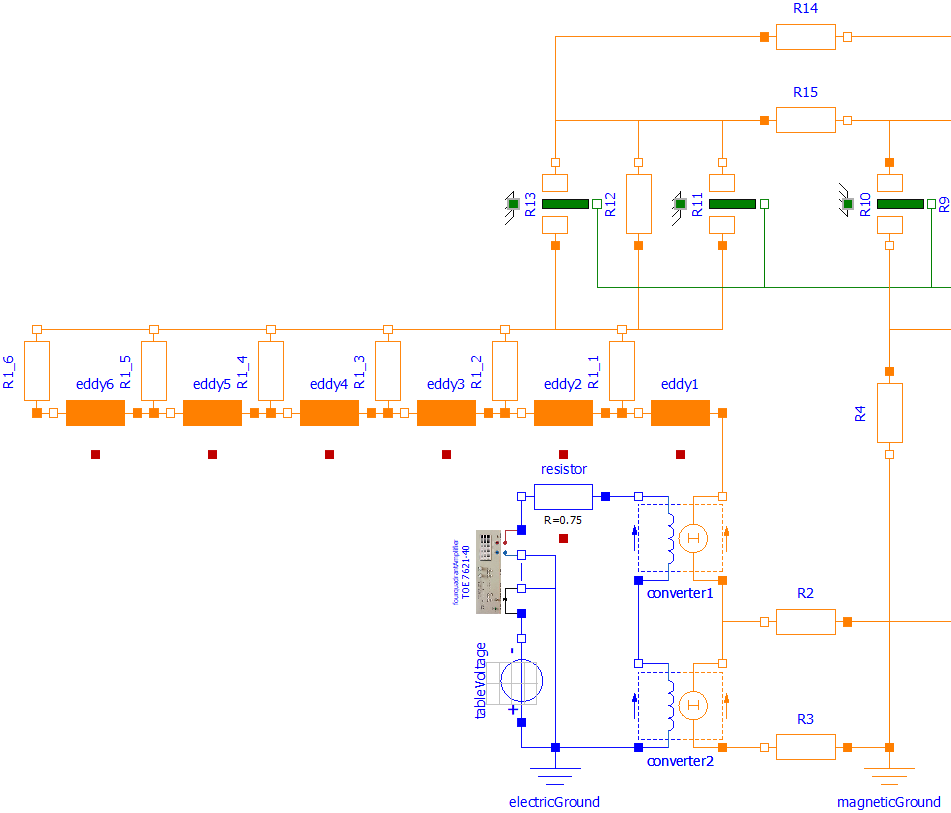}
\caption{Network model for modeling the eddy currents in the inner pole. The part not shown here (to the right)  was not altered with respect to Figure \ref{fig-17072019-1}.}
\label{fig-18032019-01}
\end{figure}

In fact, the extended range in which the delaying effect is noticeable is not spatial but temporal in nature. During the switching process, eddy currents build up in the yoke, and these eddy currents can very clearly outlast the switch-off of the excitation current. In order to investigate this quantitatively, we replace the reluctance, which in Figure \ref{fig-17072019-1} by itself represents the inner pole, in Figure \ref{fig-18032019-01} with a network of hollow cylindrical reluctances, which are coupled by damping terms \cite{Kallenbach,Stroehla}. In this way we add in Hopkinson's law:
\begin{equation*}
V_m=R_m\Phi
\end{equation*}
the eddy currents 
 induced in any of the layers of the core:
\begin{equation*}
V_m=R_m\Phi+\frac{1}{R_\text{el}}\dot{\Phi}
\end{equation*}
We then identify $L_m=1/R_\text{el}$ as magnetic inductance and can capture the eddy currents in our network using \textit{Modelica.\-Magnetic.\-Flux\-Tubes.\-Basic.\-Eddy\-Current} as shown in Figure \ref{fig-18032019-01}. Here, the flux in the inner pole was divided into six concentric regions, each with an identical cross-sectional area. The magnetic inductivities (electrical resistances of hollow cylindrical eddy current paths) result from the geometry and the experimentally determined resistivity of the X6CrMoS17 (0.774~\textmu$\Omega$\,m, see above).

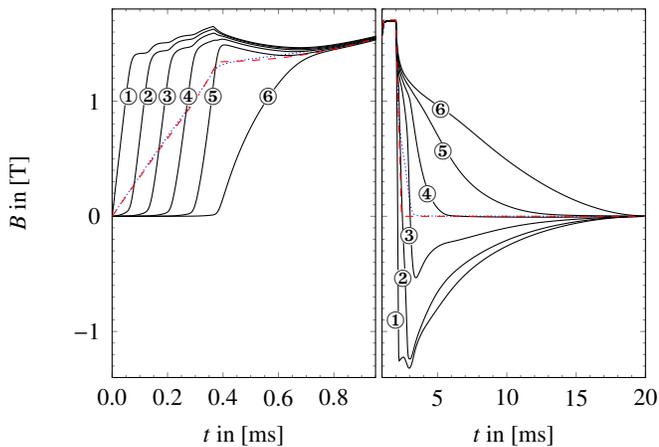
\begin{figure}
\unitlength=0.01\linewidth
\begin{picture}(100,69.5)
\put(-1.5,0){
\begin{tikzpicture}
\begin{groupplot}[group style={group size=2 by 1,
		    xlabels at=edge bottom,
               xticklabels at=edge bottom,
               yticklabels at=edge left,
               vertical sep=0.01\linewidth,
               horizontal sep=0.01\linewidth},
               height=0.75\linewidth,width=0.588\linewidth]
\nextgroupplot[
      xticklabel style={/pgf/number format/precision=1,/pgf/number format/fixed,/pgf/number format/fixed zerofill,
     /pgfplots/ticklabel shift=0.005\linewidth,/pgfplots/major tick length=0.008\linewidth,/pgfplots/minor tick length=0.004\linewidth},
     yticklabel style={/pgf/number format/precision=0,/pgf/number format/fixed,/pgf/number format/fixed zerofill,
     /pgfplots/ticklabel shift=0.005\linewidth,/pgfplots/major tick length=0.008\linewidth,/pgfplots/minor tick length=0.004\linewidth},
ylabel={$B$ in [T]},
xlabel={$t$ in [ms]},
     minor x tick num=3,
     minor y tick num=9,
     xmin=0, xmax=0.95,
     ymin=-1.4, ymax=1.8,
]
       \addplot[color=blue,densely dotted] table[x index=1,y expr=-\thisrowno{9}] {18032019_B2.txt};
       \addplot[color=black,thin] table[x index=1,y expr=-\thisrowno{3}] {18032019_B2.txt};
       \addplot[color=black,thin] table[x index=1,y expr=-\thisrowno{4}] {18032019_B2.txt};
       \addplot[color=black,thin] table[x index=1,y expr=-\thisrowno{5}] {18032019_B2.txt};
       \addplot[color=black,thin] table[x index=1,y expr=-\thisrowno{6}] {18032019_B2.txt};
       \addplot[color=black,thin] table[x index=1,y expr=-\thisrowno{7}] {18032019_B2.txt};
       \addplot[color=black,thin] table[x index=1,y expr=-\thisrowno{8}] {18032019_B2.txt};
       \addplot[color=red,dashed] table[x index=1,y expr=-\thisrowno{2}] {18032019_B2_noWirbel.txt};\nextgroupplot[
     xticklabel style={/pgf/number format/precision=0,/pgf/number format/fixed,/pgf/number format/fixed zerofill,
     /pgfplots/ticklabel shift=0.005\linewidth,/pgfplots/major tick length=0.008\linewidth,/pgfplots/minor tick length=0.004\linewidth},
     yticklabel style={/pgf/number format/precision=0,/pgf/number format/fixed,/pgf/number format/fixed zerofill,
     /pgfplots/ticklabel shift=0.005\linewidth,/pgfplots/major tick length=0.008\linewidth,/pgfplots/minor tick length=0.004\linewidth},
     xlabel={$t$ in [ms]},
     minor x tick num=4,
     minor y tick num=9,
     xmin=1, xmax=20,
     ymin=-1.4, ymax=1.8,
]
       \addplot[color=blue,densely dotted] table[x index=1,y expr=-\thisrowno{9}] {18032019_B20.txt};
       \addplot[color=black,thin] table[x index=1,y expr=-\thisrowno{3}] {18032019_B20.txt};
       \addplot[color=black,thin] table[x index=1,y expr=-\thisrowno{4}] {18032019_B20.txt};
       \addplot[color=black,thin] table[x index=1,y expr=-\thisrowno{5}] {18032019_B20.txt};
       \addplot[color=black,thin] table[x index=1,y expr=-\thisrowno{6}] {18032019_B20.txt};
       \addplot[color=black,thin] table[x index=1,y expr=-\thisrowno{7}] {18032019_B20.txt};
       \addplot[color=black,thin] table[x index=1,y expr=-\thisrowno{8}] {18032019_B20.txt};
       \addplot[color=red,dashed] table[x index=1,y expr=-\thisrowno{2}] {18032019_B20_noWirbel.txt};
\end{groupplot}
\end{tikzpicture}}
\put(9,35){
\put(10,20){{\color{white}\circle*{3}}\put(-4.35,-1.28){\ding{172}}}
\put(13,20){{\color{white}\circle*{3}}\put(-4.35,-1.28){\ding{173}}}
\put(16,20){{\color{white}\circle*{3}}\put(-4.35,-1.28){\ding{174}}}
\put(19.25,20){{\color{white}\circle*{3}}\put(-4.35,-1.28){\ding{175}}}
\put(23,20){{\color{white}\circle*{3}}\put(-4.35,-1.28){\ding{176}}}
\put(31.5,20){{\color{white}\circle*{3}}\put(-4.35,-1.28){\ding{177}}}
}
\put(47,25){
\put(13.1,-4.5){{\color{white}\circle*{3}}\put(-4.35,-1.28){\ding{172}}}
\put(14.2,2){{\color{white}\circle*{3}}\put(-4.35,-1.28){\ding{173}}}
\put(15.2,8.5){{\color{white}\circle*{3}}\put(-4.35,-1.28){\ding{174}}}
\put(18,15){{\color{white}\circle*{3}}\put(-4.35,-1.28){\ding{175}}}
\put(20.4,21.5){{\color{white}\circle*{3}}\put(-4.35,-1.28){\ding{176}}}
\put(20,28){{\color{white}\circle*{3}}\put(-4.35,-1.28){\ding{177}}}
}
\end{picture}
\caption{Field displacement in iron. Black solid lines: Flux density in hollow cylindrical flux tubes in the inner pole, marked as \ding{172} to \ding{177} from outside in. Blue dotted line: Effective flux density in inner pole. Red dashed line: Comparison with behavior without eddy currents. Switch on at 0~ms, switch off at 2~ms.}
\label{fig-14032019-01}
\end{figure}

Figure \ref{fig-14032019-01} shows the resulting flux density distribution in the inner pole during the previously considered switching process (switch-on at 0 and switch-off at 2~ms). Please note the time scale for both processes, which is displayed differently by a factor of twenty. The delayed penetration of the flux from the outside \ding{172} to the inside \ding{177} of the yoke can clearly be observed in the switch-on process. The wavy structure after reaching the nominal flux density value is an artifact of discretization into concentric hollow cylinders. The resulting effective flux density in the inner pole (blue dotted line) hardly differs from the flux density without considering the eddy currents (red dashed line). This is consistent with the observation in Figure \ref{fig-26112019-01} that the switch-on process is described very accurately without considering iron losses. The flux densitiy inside of the yoke is decidedly inhomogeneous, but the effective error to disregard this is minimal. 

In contrast, the effective flux density with eddy currents and the flux density without eddy currents clearly differ in the switch-off process. The difference in Figure \ref{fig-14032019-01} appears small only because the abscissa is scaled significantly different. The horizontal distance between the blue dotted line and the red dashed line near $B=0$ is of the order of one millisecond. This is actually the reason for the observed delay in the switch-off process: we switch off the excitation current, but we cannot switch off the eddy currents in the same way, which in turn stabilize a residual flux that holds the armature close to the yoke beyond the time expected without eddy currents.

Here too, however, the effect inside the yoke is much more varied than the examination of the effective flux density would suggest. The eddy current in the innermost element \ding{177} has essentially decayed only approx. 16~ms after switching off (near 18~ms in absolute numbers). However, this is compensated by the fact that the outermost elements \ding{172} -- \ding{174} react very quickly to the rate of change of the current linkage, resulting in a negative flux density. This consistent state is not further influenced from the outside, but decays by itself due to Joule's damping of the eddy currents. Seen from the outside, it largely balances out to an effectively vanishing flux density, but only after a cycle time of about 18~ms does it truly correspond to the initial state again. 

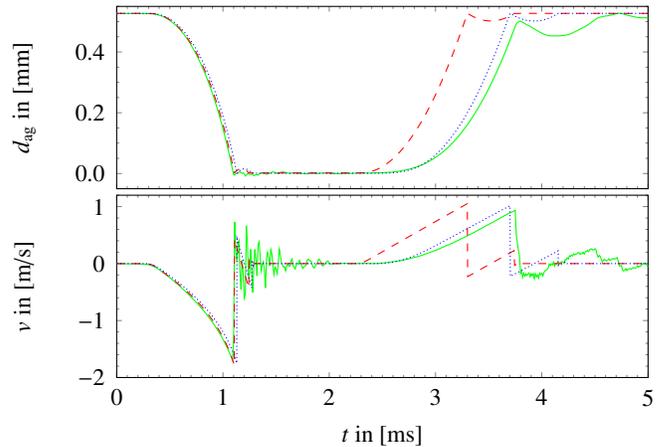
\begin{figure}
\unitlength=0.01\linewidth
\begin{picture}(100,69.5)
\put(-1.5,0){
\begin{tikzpicture}
\begin{groupplot}[group style={group size=1 by 2,
		    xlabels at=edge bottom,
               xticklabels at=edge bottom,
               yticklabels at=edge left,
               vertical sep=0.01\linewidth,
               horizontal sep=0.01\linewidth},
               height=0.4635\linewidth,width=1\linewidth]
\nextgroupplot[
      xticklabel style={/pgf/number format/precision=0,/pgf/number format/fixed,/pgf/number format/fixed zerofill,
     /pgfplots/ticklabel shift=0.005\linewidth,/pgfplots/major tick length=0.008\linewidth,/pgfplots/minor tick length=0.004\linewidth},
     yticklabel style={/pgf/number format/precision=1,/pgf/number format/fixed,/pgf/number format/fixed zerofill,
     /pgfplots/ticklabel shift=0.005\linewidth,/pgfplots/major tick length=0.008\linewidth,/pgfplots/minor tick length=0.004\linewidth},
ylabel={$d_\text{ag}$ in [mm]},
     minor x tick num=9,
     minor y tick num=3,
     xmin=0, xmax=5,
     ymin=-0.05, ymax=0.55,
]
	\addplot[color=green,thin,samples=100] table[x expr=\thisrowno{0}-10,y index=2]{Schalten_80_zu_kurz2.txt}; 
	\addplot[color=green,thin,samples=100] table[x index=0,y index=2]{Schalten_80_auf_kurz.txt}; 
	\addplot[color=red,dashed,samples=100] table[x index=0,y index=1]{Modelica_Schaltvergleich2.txt}; 
	\addplot[color=blue,densely dotted,samples=100] table[x index=1,y index=3]{18032019_dv.txt}; 
\nextgroupplot[
     xticklabel style={/pgf/number format/precision=0,/pgf/number format/fixed,/pgf/number format/fixed zerofill,
     /pgfplots/ticklabel shift=0.005\linewidth,/pgfplots/major tick length=0.008\linewidth,/pgfplots/minor tick length=0.004\linewidth},
     yticklabel style={/pgf/number format/precision=0,/pgf/number format/fixed,/pgf/number format/fixed zerofill,
     /pgfplots/ticklabel shift=0.005\linewidth,/pgfplots/major tick length=0.008\linewidth,/pgfplots/minor tick length=0.004\linewidth},
     ylabel={$v$ in [m/s]},
     xlabel={$t$ in [ms]},
     minor x tick num=9,
     minor y tick num=4,
     xmin=0, xmax=5,
     ymin=-2.0, ymax=1.2,
]
	\addplot[color=green,thin,samples=100] table[x expr=\thisrowno{0}-10,y index=1]{Schalten_80_zu_kurz2.txt}; 
	\addplot[color=green,thin,samples=100] table[x index=0,y index=1]{Schalten_80_auf_kurz.txt}; 
	\addplot[color=red,dashed,samples=100] table[x index=0,y index=2]{Modelica_Schaltvergleich2.txt}; 
	\addplot[color=blue,densely dotted,samples=100] table[x index=1,y index=4]{18032019_dv.txt}; 
\end{groupplot}
\end{tikzpicture}}
\end{picture}
\caption{Influence of eddy currents. Green solid line: measurement; blue dotted line: calculation including eddy currents; red dashed line: calculation excluding eddy currents. 
Top: armature stroke, bottom: armature speed.}
\label{fig-15032019-01}
\end{figure}

Figure \ref{fig-15032019-01} shows the effect of these eddy currents on the dynamic behaviour of the system, together with the experimental data and the simulation without eddy currents as shown previously in Figure \ref{fig-26112019-01}. All three agree in the switch-on behavior; here there had been no problem from the beginning. By modelling the eddy currents, however, we now also get a very reasonable agreement in the switch-off behaviour. While we essentially had a free parameter for the hysteresis effects with the minute fine tuning of the residual air gap, which allowed us to adjust the displacement of the curve arbitrarily, yet not meaningfully, we have no such parameter here. The effective, simulated delay results directly from the material properties of the yoke and the geometry. The observed agreement can therefore be interpreted as an indicator that the selected model is at least qualitatively accurate.

\section{Conclusion}
We have investigated the dynamic behaviour of a switching magnet. Experimentally, the armature movement was determined contactlessly by a laser vibrometer. The starting point of the simulation was a network whose validity in the magnetostatic case had already been confirmed. Two different aspects of iron losses were investigated. Firstly, hysteresis effects, especially remanent magnetization after switch-off of the current linkage. It was found that for the soft iron used, these influences are limited to the first few micrometers of residual air gap (starting from the ideal closed circuit), and their amount depends sensitively on the actual residual air gap. This is inconsistent with the experimentally robust observation of a delayed switch-off process. We secondly considered eddy current effects in the inner pole. This does not only provide results for the armature dynamics that are consistent with the experimental findings, but also provides insights into the internal flux and eddy current distribution during dynamic operation of the electromagnet.  

\section*{Acknowledgments}
Herbert Schmidt would like to thank J$\ddot{\text{o}}$rg Frochte for many fruitful discussions. The authors would like to thank Detlef Herkt and his team for manufacturing the functional model. 

\bibliography{HSchmidt-Modelica2020_arXiv}

\end{document}